\documentclass[]{rspublic}

\usepackage{natbib}
\usepackage{graphicx}
\usepackage{amsmath}
\usepackage{amssymb}

\title[Models for Gamma-Ray Bursts and Diverse Transients]{Models for
  GRBs and diverse transients} 

\author[S. E. Woosley and W. Zhang]{S. E. Woosley$^{1}$ and
    Weiqun Zhang$^{2}$}

\affiliation{$^{1}$Department of Astronomy and
    Astrophysics, UCSC, Santa Cruz CA 95064, USA\\ $^{2}$Kavli
    Institute for Particle Astrophysics and Cosmology, Stanford
    University, Stanford CA 94309, USA}

\def\Msun {M$_{\scriptscriptstyle \odot}$}
\def \ltaprx {\lower .1ex\hbox{\rlap{\raise .6ex\hbox{\hskip .3ex
	{\ifmmode{\scriptscriptstyle <}\else 
		{$\scriptscriptstyle <$}\fi}}}
	\kern -.4ex{\ifmmode{\scriptscriptstyle \sim}\else 
		{$\scriptscriptstyle\sim$}\fi}}}
\def\gtaprx {\lower .1ex\hbox{\rlap{\raise .6ex\hbox{\hskip .3ex
	{\ifmmode{\scriptscriptstyle >}\else 
		{$\scriptscriptstyle >$}\fi}}}
	\kern -.4ex{\ifmmode{\scriptscriptstyle \sim}\else 
		{$\scriptscriptstyle\sim$}\fi}}}

\begin{document}

\date{}


\maketitle

\label{firstpage}

\begin{abstract}{gamma-ray burst -- models}
The observational diversity of ``gamma-ray bursts'' (GRBs) has been
increasing, and the natural inclination is a proliferation of
models. We explore the possibility that at least part of this
diversity is a consequence of a single basic model for the central
engine operating in a massive star of variable mass, differential
rotation rate, and mass loss rate. Whatever that central engine may be
- and here the collapsar is used as a reference point - it must be
capable of generating both a narrowly collimated, highly relativistic
jet to make the GRB, and a wide angle, sub-relativistic outflow
responsible for exploding the star and making the supernova bright.
To some extent, the two components may vary independently, so it is
possible to produce a variety of jet energies and supernova
luminosities. We explore, in particular, the production of low energy
bursts and find a lower limit, $\sim$10$^{48}$ erg s$^{-1}$ to the
power required for a jet to escape a massive star before that star
either explodes or is accreted. Lower energy bursts and and
``suffocated'' bursts may be particularly prevalent when the
metallicity is high, i.e., in the modern universe at low redshift.
\end{abstract}

\section{Introduction}

If the BATSE era was the age of discovery for GRBs, and the
BeppoSax/HETE era, the age of cosmology (or at least when we clearly
saw that bursts were at cosmological distances and associated with
massive stars), then SWIFT may be remembered, in part, as the age of
increasing diversity. GRBs now come in short and long varieties, as
well as hybrids having properties of both long-soft and short hard
bursts. There are long-soft cosmological x-ray transients, GRBs with
supernovae, GRBs without supernovae, and energetic supernovae with
weak GRBs. While it would be surprising if all cosmolgical transients
lasting within 0.1 to 1000 s, with a power spectrum peaking between
1 keV and 1 MeV were the same thing, the parameter space for massive
stars that die and produce a rapidly rotating compact remnant is
really quite large, and diverse transients are to be expected.

\section{A Two-Component Model}

\begin{figure}
\begin{center}
\includegraphics[width=84mm]{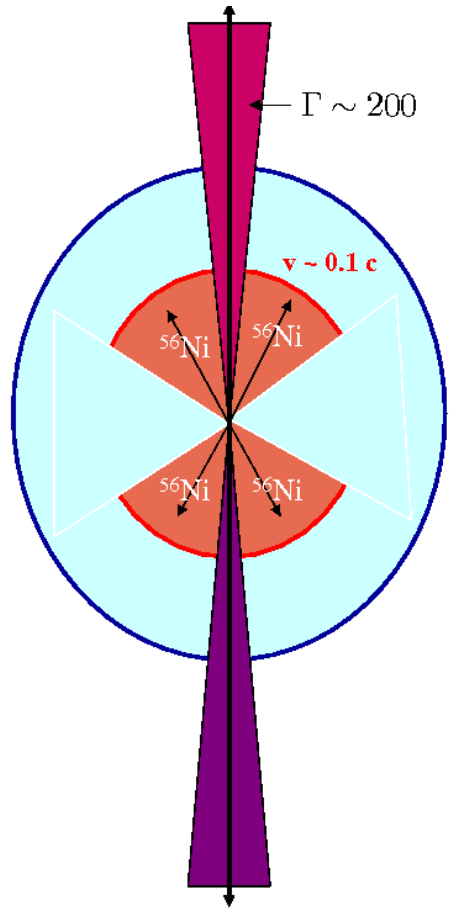}
\caption{The jet in those GRBs which have accompanying supernovae must
  have at least two components - a narrow highly relativistic jet
  responsible for the burst itself and a broad subrelativistic outflow
  responsible for exploding the star and producing the $^{56}$Ni to
  make it bright. The broad outflow extends to at least ~ 1
  radian. There may additionally be a mildly relativistic outflow (not
  shown) from the cocoon explosion that contributes to the afterglow
  and off-axis bursts.}
\end{center}
\end{figure}

\subsection{The Need for Two Components}

Any GRB that is accompanied by a stellar explosion with energy
$\gtaprx$10$^{51}$ erg must consist, inside the star that makes it, of
at least two outflows - a highly relativistic core focused to a tiny
fraction of the sky (typically a few tenths of a percent), and a broad
angle, subrelativistic outflow (Fig. 1). The core jet, by itself, will
not violently explode the star (though the jet may be surrounded by a
powerful cocoon), and no isotropic explosion with credible energy will
give adequately relativistic ejecta to make a common GRB.  In the
millisecond magnetar model, the large-angle component could be an
``ordinary'' supernova, launched by neutrinos or MHD processes, and
the narrow jet might be an afterthought, an MHD collimated outflow
that happens after the supernova shock has already been launched. In
the collapsar model, the jet is produced by MHD processes near the
black hole, while the large angle outflow is from the disk wind
\citep{Mac03}.  It is the large angle outflow that is responsible both
for most of the supernova's kinetic energy and the $^{56}$Ni needed to
make it bright.

\subsection{Variable Supernova Brightness}

Because these two components have different origins, they are free, to
some extent, to vary independently. In the collapsar model, whatever
makes the jet involves physics very near the black hole. In the
\citet{Bla77} model, for example, it is the rotation of the hole
itself that powers the jet, the disk playing a passive role. In the
neutrino version of the collapsar model \citep{Mac99}, the jet is
powered by neutrino annihilation along the rotational axis while the
broad angle outflow comes from the ``disk wind'', energized by viscous
processes farther out \citep{Mac99,Nar01}. It is not difficult to
envision situations where the relative importance of these two
components varies.  If the collapsing star has less angular momentum
in its core, which is expected to be the case, for example, when the
star has lost a lot of mass along the way, it will make a larger black
hole before forming a disk. The accretion rate from the collapse of
the lower density mantle farther out will also be less. It may be
difficult in these conditions to realize a stable, neutrino-dominated
accretion disk \citep{Nar01}.  The accretion may be oscillatory
\citep{Woo06c} and the composition of the wind - if there is one - is
unknown. It may be some other nucleosynthetic product than $^{56}$Ni

\section{The Effects of Metallicity}

The importance of metallicity in producing a GRB was pointed out by
\citet{Mac99} and several observational studies have suggested a
correlation of GRBs with low metallicity regions
\citep[e.g.,][]{Fru06}, though see \citet{Pro06}.

In theory, metallicity and the final rotation rate of the stellar core
are inversely correlated. This is because mass loss is dependent upon
the iron abundance, especially during the critical Wolf-Rayet stage of
the progenitor evolution that precedes the burst
\citep{Vin05,Woo06b,Yoo05,Yoo06}. Mass loss saps angular momentum from
the surface and the matter that expands to take the place of the lost
matter rotates more slowly.  This slower rotation is communicated by
torques and circulatory currents to the layers deeper inside. Without
an angular momentum, $j \sim 3 \times 10^{15}$ cm$^2$ s$^{-1}$, the
rotational energy of a neutron star would be too low to power a GRB,
let alone an energetic supernova like SN 1998bw. To form a disk around
a 3 \Msun \ black hole in the collapsar model requires at least $2
\times 10^{16}$ cm$^2$ s$^{-1}$. With less rotation, a supernova might
still be possible, and even some sort of low energy transient, but for
$j \ltaprx 10^{15}$ cm$^2$ s$^{-1}$, some other means besides rotation
must be found to blow up the star.

For those stars that do die with adequate angular momentum, the
properties of the GRBs that they make should be correlated with the
excess above these minimum values. Stars that lose less mass, on the
average, die with more rapid rotation rates {\sl and} greater
mass. The latter characteristic makes them more tightly bound and
gives greater accretion rates on the young collapsed core, making
black hole formation more likely \citep{Fry99}. Both these properties
favor the formation of a collapsar. Stars with the greater angular
momentum form their disks earlier around less massive black holes. The
temperature at the last stable orbit is then higher \citep{Pop99},
which increases neutrino emission and allows the disk to dissipate its
energy more efficiently. These neutrinos annihilate in a smaller
volume making a neutrino-powered jet feasible. The total amount of
mass available for accretion is also greater which makes for a longer
more energetic burst.  These properties make it more likely that the
most energetic gamma-ray bursts will occur in regions with the lowest
metallicity, though there can be considerable variation in individual
events because of the mass of the star itself varies.

For higher metallicity, and hence less rotation, the black hole in the
collapsar model forms later and accretes more slowly. The dynamical
time scale for the matter that is left outside is longer, but there is
also less of it. Neutrino emission becomes less effective, both as a
power source for the burst and a means of dissipating disk binding
energy. On the average, these bursts will last longer and have less
total energy. 

This all assumes a one-to-one relation between mass that falls to the
center and mass that accretes, but even when a stable disk forms, that
disk may experience considerable loss to a wind
\citep{Mac99,Koh05}. In fact, it is this wind, not the jet, that is
responsible for blowing up the star. Either a low accretion rate (low
rotation, high metallicity) or a very high angular momentum (high
rotation, low metallicity) can reduce the neutrino losses and make
black hole accretion less efficient. In the former case, the low
accretion rate makes the temperature too low for effective neutrino
dissipation. In the latter, the disk forms at such a large radius that
neutrino dissipation is ineffective. If one associates the disk wind
with the brightness of the supernova and black hole accretion with the
strength of the GRB, then it is clear that considerable variation in
the ratio can be expected.

For still lower rotation rates, making a GRB using black hole
accretion becomes increasingly difficult and, if there is to be an
energetic jet at all, one needs to consider neutron star models. In
fact, the neutron star possibility is there all along, provided
something (neutrinos?) can hold up the accreting star while the
``proto-neutron star'' experiences several seconds of Kelvin-Helmholtz
evolution and shrinks to its final radius, rotation rate, and magnetic
field strength \citep{Tho04,Woo06a}. To explain GRBs in the highest
metallicity regions (i.e., solar), and lightest presupernova
progenitors, it may be that this mechanism is necessary. On the other
hand, neither the energy of a GRB nor its duration is a unique
signature of a neutron star model, and there may be other paths to
making a GRB in metal-rich regions that involve binary stars

\begin{figure}
\begin{center}
\includegraphics[width=84mm]{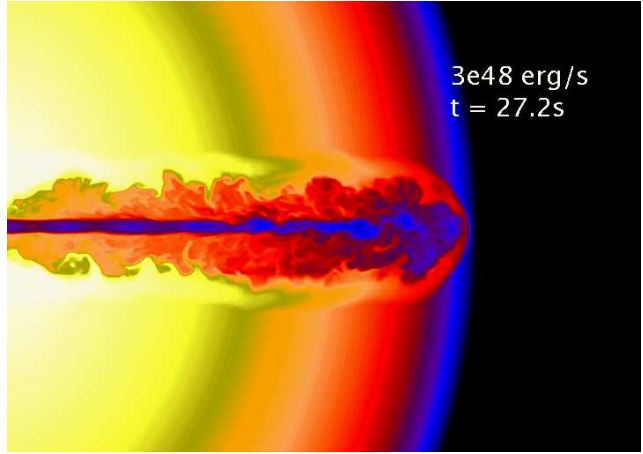}
\caption{Three-dimensional calculation of a relativistic jet of $3
\times 10^{48}$ erg s$^{-1}$ introduced at $1 \times 10^{10}$ in a 15
\Msun \ Wolf-Rayet presupernova star of radius $8 \times 10^{10}$
cm. The initial jet had Lorentz factor 5, total energy to mass ratio
40 and an initial cylindrical radius $1 \times 10^9$ cm ($\sim$5
degrees). Plotted is the logarithm of the density as the jet nears the
surface. The jet took much longer to reach the surface than a similar
jet with power $3 \times 10^{50}$ erg s$^{-1}$ studied by
\citet{Zha04} and was less stable.  After break out, the jet
eventually becomes more stable as an opening is cleared by the
relativistic flow. For greater detail see \citet{Zha07}.}
\end{center}
\end{figure}

\begin{figure}
\begin{center}
\includegraphics[width=84mm]{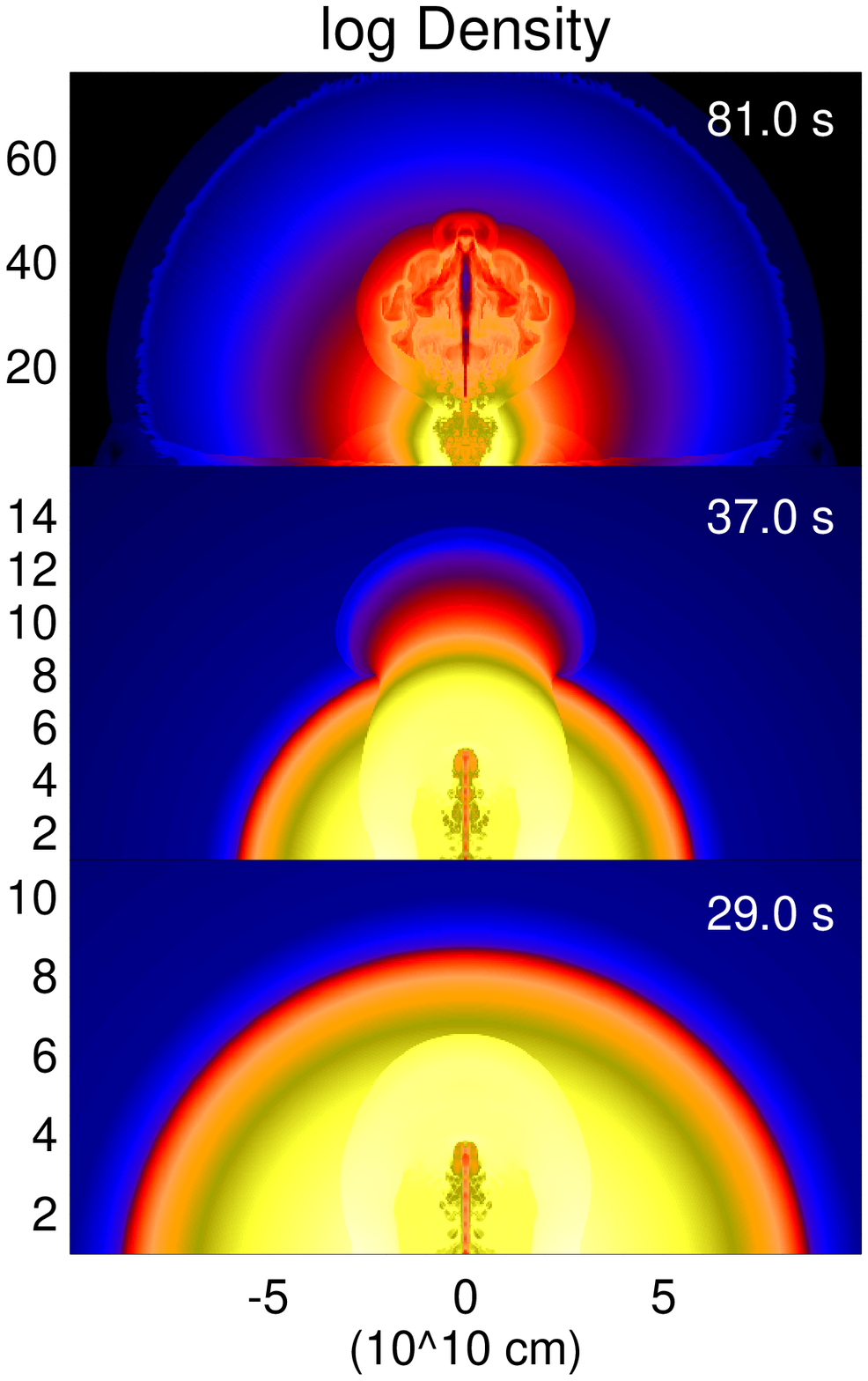}
\caption{A still lower energy jet ($1.0 \times 10^{48}$ erg s$^{-1}$)
is launched inside a star that is already in the process of exploding
as a supernova (10$^{51}$ erg deposited nearly isotropically within a
cylindrical radius of $1.7 \times 10^{10}$ cm near the origin at the
bottom). The plot shows the density. The outer boundary of the
supernova shock is visible at 29 and 37 s as the edge of the
yellow-white oval. The jet power, $1 \times 10^{48}$ erg s$^{-1}$, is
so low that the it takes a very long time to break out. Still, if the
power can be maintained for hundreds of seconds a bright transient
could still be observed.At 81 s, the jet has expanded to 5 times the
initial radius of the star and has overtaken the supernova shock.}
\end{center}
\end{figure}

\section{Low Energy Bursts}

\subsection{A Minimum Power to Break Out}

Whether produced by a neutron star or by black hole accretion, the
power of the jet in massive star models for GRBs is likely to vary
significantly from source to source. In the collapsar model, the power
is sensitive to the accretion rate, the stability of the accretion
disk, the efficiency for turning accretion energy into directed
relativistic outflows, and the rotation rate and mass of the black
hole. In the millisecond magnetar model it is sensitive to the
rotation rate and field of the neutron star and unknown efficiencies
for producing collimated outflows. It is thus reasonable to inquire as
the the outcome of jets of various powers released deep within the
progenitor star. How long does it take jets of varying energy to break
out? Is there a minimum power below which jets do not make it out at
all, or takes such a long time that the star would surely have
collapsed or exploded in the meantime? This would imply an interesting
limit on the power of bursts that can be produced directly by internal
shocks within the jet. It is not a stringent lower bound, however,
because the fraction of the jet energy in highly relativistic matter,
and especially the efficiency for converting kinetic energy into
gamma-rays is also uncertain and possibly highly variable.

Three jets of varying power, 0.03, 0.3, and $3 \times 10^{50}$ erg
s$^{-1}$, were introduced at 10$^{10}$ cm in a 14 \Msun\ Wolf-Rayet
star of radius $8 \times 10^{10}$ cm.  The mass interior to $10^{10}$
cm was removed from the presupernova star and replaced by a point
mass.  Each jet is defined by its power, initial Lorentz factor (here
$\Gamma$ = 5), and the ratio of its total energy (excluding rest mass
energy) to its kinetic energy (here 40). The parameters here are such
that if the jet expanded freely to infinity, its Lorentz factor would
be $\Gamma$ = 200.  The grid adopted in the 3D study
\cite{Zha03,Zha04} is Cartesian with 256 zones each along the $x$- and
$y$-axes and 512 along the $z$-axis (jet axis). For full details of
the calculations see \citet{Zha07}.

Fig. 2 shows the density structure just as the $3 \times 10^{48}$ erg
s$^{-1}$ jet erupts from the surface of the star 27 s after initiation
at 10$^{10}$ cm. The two other higher energy jets, 0.3 and $3 \times
10^{50}$ erg s$^{-1}$ took 15 s and 7 s respectively. It is also
apparent in Fig. 2 that the structure of the jet itself is less
coherent at low energy \citep[compare with][]{Zha04}.

The long time for the jet to break out in Fig. 2 is comparable to the
collapse time of the core, the sound crossing time for the core, and
the duration of a common GRB. An attempt to push a jet of only
10$^{48}$ erg s$^{-1}$ ended in failure. The jet had not emerged after
100 s.  While a relativistic jet of arbitrarily low power will
eventually break out of any star, the star in this case would either
have largely accreted, reducing the energy of the jet further, or
blown up. If it blew up, then the jet would have still further to go
before breaking out.

If one assumes, as seems reasonable, that jets are unlikely to produce
bursts much shorter than the time it takes them to escape their
progenitor star, these calculations suggest that lower energy bursts
will last longer. Ultra-relativistic jets with angle-integrated power
much less than 10$^{48}$ erg s$^{-1}$ may be hard to make in massive
stars. The energy the jet deposits in the star on the way out,
essentially the break-out time times the power, also declines as the
jet power is turned down. Thus, without the broad angle component (not
included here), the supernovae accompanying weaker GRBs would also be
weaker \cite{Zha07}. Including the broad component can change this
radically ($\S$2).

\begin{figure}
\begin{center}
\includegraphics[width=84mm]{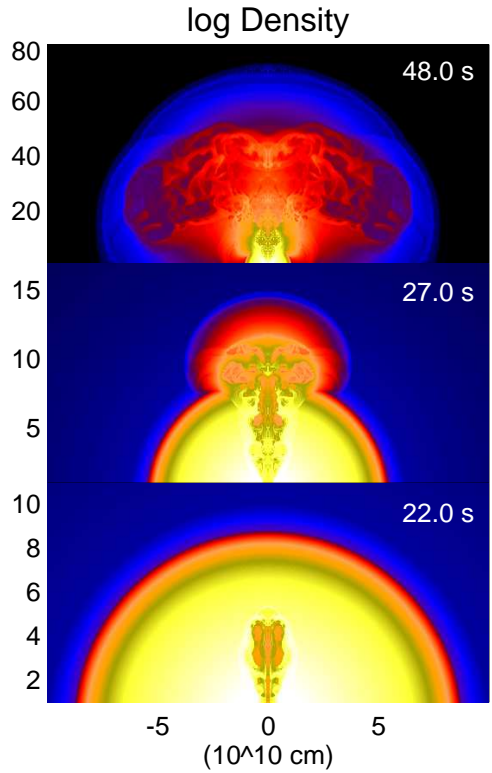}
\caption{A jet of $3 \times 10^{49}$ erg s$^{-1}$ is turned off just
before break out to simulate the cessation of accretion at the center
of the star. The jet still caused a very asymmetric explosion and
ejected mildly relativistic matter, but no hyper-relativistic jet
core. There would be no GRB by internal shocks. However, there is
mildly relativistic ejecta (Fig. 5), and the interaction of that matter
with the presupernova wind would produce an energetic transient
\citep{Tan01,Zha07}.}
\end{center}
\end{figure}

\begin{figure}
\begin{center}
 \includegraphics[width=84mm]{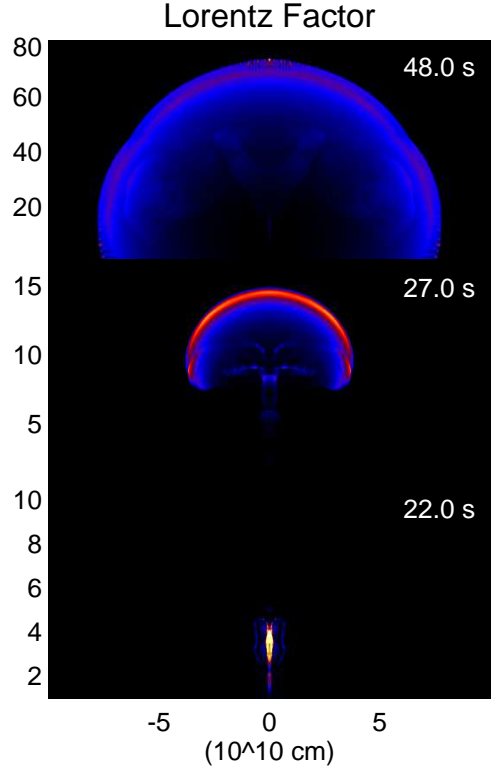}
 \caption{Lorentz factor corresponding to the models in Fig. 4. Red
and purple correspond to Lorentz factors of about 3 and 2
respectively.  At 27 s in this run, there are $1.5 \times 10^{-8}$
\Msun \ and $1.2 \times 10^{47}$ erg in material with a Lorentz factor
greater than 2 and no material that has a Lorentz factor greater than
5.}
\end{center}
\end{figure}

\subsection{Decaying Jets}

A jet of lower lower takes longer to break out, but then the longer it
takes, the more likely it is that the jet source has decreased further
in power or the star has blown up already or both. Either case leads
to a situation where a jet of declining energy finds itself still
deeply embedded in dense, very optically thick matter.

Fig. 3 shows an event where a supernova with kinetic energy
$\sim10^{51}$ erg has already happened and its initial blast reached
the surface of the star before a weak jet, in this case 10$^{48}$ erg
s$^{-1}$, finally arrives. The total energy in this jet, about
10$^{50}$ erg per jet (and there is one at the other pole), is not
especially small, just its power. A jet of any power will eventually
break out, provided it is artificially maintained long enough, but
unless it does there is no GRB. At both 37 and 81 s in Fig. 3 there is
no material with $\Gamma$ greater than about 2 except what is inside
the jet, i.e., the supernova itself makes no strong hard transient. If
the jet suddenly lost power at its base before 80 s, it would quickly
share its energy with a large mass, and there would be no major
relativistic event.

This point is perhaps more clearly made in the model shown in Fig. 4
and 5.  Here no supernova was assumed, but the jet, with power $3
\times 10^{49}$ erg s$^{-1}$, was abruptly turned off at the origin at
22 s, just a few seconds before it would have erupted from the surface
of the star.  A highly asymmetric explosion still resulted, but the
Lorentz factor of the jet, which would have been 200 had it coasted
unimpeded to infinity, was quickly braked by running into the matter
above. The blast spread sideways and by 48 s had become essentially
isotropic so far as the relativistic matter was concerned. The total
relativistic energy, $\Gamma$ greater than 2, here was only 10$^{47}$
erg, concentrated in about 10$^{-8}$ \Msun. For a mass loss rate of
10$^{-5}$ \Msun \ y$^{-1}$ before the explosion, this bit of matter
would decelerate at a few $\times 10^{12}$ cm in a region which is
optically thin. A transient lasting a few 10's of seconds, somewhat
like GRB 980425 \citep{Woo99,Tan01}, might result, although the
total amount of relativistic matter in 980425 was probably at least
two orders of magnitude greater.

However, the case shown in Fig. 4 is extreme. It is not natural that
the central engine abruptly turns off at some point. The decline
should be gradual. This would be the case for a steadily slowing
pulsar at the origin, or for a black hole with a declining accretion
rate. The results shown in Fig. 4 and 5 would then be augmented by
additional energy and a higher Lorentz factor component that would be
increasingly centrally concentrated at late times. A change of less
than a factor of two in the total jet energy could affect the energy
in relativistic matter by orders of magnitude, provided that just a
little of that energy is provided to the jet after it has broken free.

\subsection{Enshrouded Bursts}

Finally, any jet that is to produce a burst visible from far away must
not only escape the star or supernova, but emerge intact from any
optically thick wind near the star. The density in the wind,
neglecting clumping and time variability, is about $\rho \sim 5 \times
10^{-13} \ \dot M_{-5} \ r_{12}^{-2}$ g cm$^{-3}$. The optical depth
to electron scattering from radius $r_{12} \times 10^{12}$ cm is then
$\tau \sim 0.1 \dot M_{-5}/r_{12}$. Here $\dot M_{-5}$ is the mass
loss rate in 10$^{-5}$ \Msun \ y$^{-1}$.

The relativistic jets in typical GRBs carry an equivalent isotropic
energy that is at least the equivalent of the gamma-rays they
ultimately produce. For reasonable radiative efficiencies, this
implies a jet energy $\sim$10$^{51}$ erg. With a Lorentz factor of
200, this is a rest mass of only $\sim$ a few $\times 10^{-6}$ \Msun.
Material moving at this speed would give up its energy if it
encountered $1/\Gamma$ times its rest mass, or about 10$^{-7}$
\Msun. This is to be compared with the mass loss rate, which for
typical bursts, especially in loss metallicity regions, is less than
10$^{-5}$ \Msun \ y$^{-1}$, times the solid angle of the jet, say
1\%. That is, the jet coasts to the distance the mass loss would go in
one year, about $3 \times 10^{15}$ cm before giving up all its
energy. The GRB is produced well inside that radius and the afterglow
near that radius and outside.

Consider, however, the circumstances for a jet with roughly 100 times
less energy, i.e., 10$^{49}$ erg, running into a wind with density ten
times as great, i.e., 10$^{-4}$ \Msun \ y$^{-1}$. Now, the jet
encounters $1/\Gamma$ times its mass in the wind of only the last
few hours. This corresponds to a radius of $\sim$10$^{12}$ cm and a
light crossing time of just a hundred seconds. The wind is optically
thick at the radius where this interaction occurs.  When such a
ballistic jet enters this region, it will be braked and slowed,
sharing its energy with a large mass. The explosion then becomes
nearly isotropic, with mildly relativistic matter ejected at large
radius (Fig. 4). One possibility is a long soft thermal x-ray
transient \citep{Cam06}.

\section{Conclusions}

GRBs are a rare branch of massive stellar death characterized by very
rapid, highly differential rotation. GRBs will be easier to make, and
may have more energy in stars with a lower iron abundance, for
example, at higher redshift or in dwarf galaxies. It is important to
note the key role played by iron here, not just total metallicity,
i.e., oxygen. Iron and oxygen have different histories since the
former is made mostly in Type Ia supernovae and the latter in Type
II. Iron might be more deficient compared to solar than oxygen in some
galaxies or regions of galaxies. Currently, models based upon complete
mixing on the main sequence in very rapidly rotating single stars
\cite{Woo06b,Yoo06} give an upper limit to the iron abundance allowed
in a successful collapsar model of about 30\% solar, but it is the
rate at which angular momentum is lost in the Wolf-Rayet stage that
matters, not the iron abundance itself. Less angular momentum will be
lost per gram of mass lost if the winds are strongly concentrated at
the rotational axis \cite{Mae02}. The estimates of magnetic torques
from \citet{Spr02} are highly uncertain. So too are the mass loss
rates themselves, even for solar metallicity Wolf-Rayet
stars. Therefore, single star progenitors of solar metallicity cannot
be ruled out at this point solely upon the basis of theory.

A great variety of transients are possible depending upon the power
and duration of the jet produced when the star dies, its Lorentz
factor (and the time modulation, $\Gamma$(t)), the angle at which the
event is observed, the mass loss rate, and the relative strength and
composition of the broad angle component. Each of these, except the
random viewing angle, may vary with the mass, metallicity, and
rotation rate of the massive stellar population, and hence with red
shift. On the average, one expects greater mass loss and hence slower
rotation at higher metallicity so the GRBs in the modern universe may
be qualitatively different from those long ago. In particular, they
might have lower average energy and be more affected by a
higher-density circumstellar environment.

GRBs are frequently, perhaps universally accompanied by bright Type
Ic-BL supernovae \citep{Woo06a}. Indeed, it is difficult to imagine
the production of a relativistic jet in a massive star by any means
that does not require the star's death and at least partial
disruption, and the evidence linking most long-soft GRBs to massive
stars is strong.  But the {\sl brightness} of a supernova of Type I
(the ``super'' in the ``supernova'') depends upon how much $^{56}$Ni
is made. In a collapsar it is possible that the dominant constituent
of the disk wind is not radioactive. This happens if the density where
the wind originates is unusually high, either because the viscosity is
low or the wind dominantly comes from the inner disk
\citep{Pru03,Pru04}. In fact, material with electron mole number $Y_e$
less than 0.482, will be free of $^{56}$Ni.  It is also possible that
the wind in the collapsar model, or broad angle component in the
neutron star model may, for some reason, be weak or fall back during
the explosion. Much is still to be learned about these winds and
limits on supernovae in nearby GRBs will be an important constraint.
It is unfortunate that such information on bursts at higher redshifts
- which might possibly be different beasts - is so difficult to
obtain.

Finally, there is a minimum jet power, around 10$^{48}$ erg s$^{-1}$
that is capable of escaping a massive stars in a time comparable to
the duration of most long-soft GRBs. On a longer time scale the star
will either have exploded or accreted onto its compact remnant, either
of which will affect the properties of the burst. If the star has
already exploded (e.g., by neutrinos, the disk wind, or a large angle
energy input by a pulsar), then the jet may have to play ``catch-up''
with the ejecta (Fig. 3). If the jet turns off any time in this
period, a very weak transient will result. Similarly, if the central
engine turns off before the jet breaks out (Fig. 4 and 5), no GRB
results though the explosion is still quite anisotropic and some sort
of x-ray transient may be observed. In what be the most natural case,
a jet that does eventually break out but with greatly diminished power, 
what is seen will depend greatly on viewing angle and how long the 
jet stayed on after it broke out. 

It is generally assumed that the interaction of the jet with the
external medium will be negligible during the burst phase, at least in
the internal shock model, yet dominant in the afterglow phase. For
sufficiently high mass loss rates and GRB jets with sufficiently low
energy will dissipate their energy close enough to the progenitor star
to affect the burst. In some cases where a significant part of the jet
energy has piled up in geometrically thin ``plug'' which the jet is
pushing along, deceleration in a dense medium might give a short burst
preceding a longer component from the usual internal shock interaction
\citep{Zha04}.  This might result in a ``hybrid burst'' with a brief
intense initial spike from external shock interaction followed by a
longer, more typical GRB \citep[e.g., GRB 0600614][]{Geh06}. If the jet
has less energy and dissipates within an optically thick region of the
wind, its energy may spread out to larger angles in a way analogous to
Fig. 4. One then might get a long x-ray flash visible at large
angles. An example might be XRF 060218 \citep{Cam06}.

\section*{Acknowledgments}
 
This work has been supported by the NASA (NNG05GG08G) and by the
SciDAC Program of the DOE (DE-FC02-06ER41438). Calculations were
carried out on the Columbia computer at NASA-Ames.  W.Z. has been
supported by NASA through Chandra Postdoctoral Fellowship PF4-50036
awarded by the Chandra X-Ray Observatory Center.

\label{lastpage}

\end{document}